\documentclass[fleqn,10pt]{SelfArx} 

\usepackage[english]{babel} 
\usepackage[final]{pdfpages}

\usepackage[T1]{fontenc}
\usepackage[utf8]{inputenc}
\usepackage{authblk}

\usepackage{graphicx,epsfig}

\usepackage{times}
\usepackage{graphics,dcolumn,bm,fleqn,float}
\usepackage{amssymb,amsmath,multirow,rotate,color}
\usepackage{comment} 
\usepackage{cite} 

\usepackage{graphicx,epsfig}
\usepackage{times}
\usepackage{graphics,dcolumn,bm,fleqn,float}
\usepackage{amssymb,amsmath,multirow,rotate,color}
\usepackage{comment} 

\usepackage{color}

\definecolor{color1}{RGB}{0,0,90} 
\definecolor{color2}{RGB}{0,20,20} 

\usepackage{hyperref} 
\hypersetup{hidelinks,colorlinks,breaklinks=true,urlcolor=color2,citecolor=color1,linkcolor=color1,bookmarksopen=false,pdftitle={Title},pdfauthor={Author}}

\JournalInfo{~} 
\Archive{} 

\PaperTitle{Multilayer Networks in a Nutshell} 

\Authors{\normalsize Alberto Aleta\textsuperscript{1,2}, Yamir Moreno\textsuperscript{1,2,3}}

\affiliation{\textsuperscript{1}\textit{Institute for Biocomputation and Physics of Complex Systems (BIFI), University of Zaragoza, Spain.}}
\affiliation{\textsuperscript{2}\textit{Department of Theoretical Physics, Faculty of Sciences, University of Zaragoza, Spain.}}
\affiliation{\textsuperscript{3}\textit{ISI Foundation, Turin, Italy.}}

\newcommand{\keywordname}{Keywords} 

\Abstract{Complex systems are characterized by many interacting units that give rise to emergent behavior. A particularly advantageous way to study these systems is through the analysis of the networks that encode the interactions among the system's constituents. During the last two decades, network science has provided many insights in natural, social, biological and technological systems. However, real systems are more often than not interconnected, with many interdependencies that are not properly captured by single layer networks. To account for this source of complexity, a more general framework, in which different networks evolve or interact with each other, is needed. These are known as multilayer networks. Here we provide an overview of the basic methodology used to describe multilayer systems as well as of some representative dynamical processes that take place on top of them. We round off the review with a summary of several applications in diverse fields of science.}

\begin{document}

\flushbottom 

\maketitle 

\thispagestyle{empty} 


\section{Introduction}
In 1972 Philip W. Anderson published his famous article {\it More is different} where he criticized the reductionist hypothesis according to which everything, from matter to life, could be completely described from the same set of fundamental laws \cite{Anderson1972}. The fallacy, he said, was that a reductionist approach did not imply a constructionist one. That is, being able to reduce something to its fundamental laws does not imply the ability to start from them and reconstruct it. This simple, yet powerful, idea is the key concept behind what we call today complex systems.

Complex systems are defined as those systems exhibiting emergent phenomena. The definition of emergence is, however, still a subject of intense debate. Given the scope of this review, we will settle for one of the most common definitions: emergence is related to phenomena that cannot be explained or predicted just from the constituent parts of the system (nevertheless, we refer the interested reader to Morrison et al. \cite{Morrison2015} for more information). This definition of emergence implies a hierarchical view of the world which calls again into question the role of reduction. 

A simple example of the limitations of a reductionist approach is water, as its properties cannot be predicted from those of hydrogen and oxygen on their own \cite{Mack2001}. If we do not consider them isolated but allow them to interact, the atoms will establish chemical bonds to form molecules from whose interactions we can extract most properties of water. Thus, to correctly address some systems a more holistic view is needed, in which we account for both its constituents and their interactions.

The main framework used to encode these characteristics of systems in a mathematically tractable object is graph theory. The application of statistical physics to graphs was initially aimed at studying problems like percolation processes. However, as the interest on complex systems increased, it was promptly discovered that with the addition of techniques borrowed from other scientific disciplines, as computer science or statistical inference, it was possible to go beyond classical ``physics systems'' and study the behavior of a much broader array of problems in fields as diverse as economic, social and biological sciences. The combination of these techniques provided us with a very powerful tool to study complex systems which is usually known as network science.

In its most basic form, a network reduces a system to an abstract structure consisting of some simplified entities, called nodes, which retain some of the properties of the original components of the system, and the connection patterns between them, their links. This representation not only allows for a more complete view of the system but also uncovers some novel characteristics. In particular, the structure of the networks, that is, the particular pattern of interactions of its components, can have a big effect on the behavior of the system \cite{Newman2010}.

Nonetheless, this mathematical representation might still be reduction of the original system. Thus, one may wonder whether we could incur in the same reductionism that we wanted  to avoid, albeit at a different level of description. Network based approaches have produced significant insights into the structure and dynamics of complex systems during the last two decades or so. However, it has recently been shown that it is possible to add more information into these models while keeping them tractable and making them more powerful at the

\onecolumn

\noindent same time. The way of doing it is by classifying the interactions found in a system into groups according to their characteristics. This classification yields a set of networks, one for each type of interaction, connected to each other. The way in which these networks are connected to each other, the entities that nodes represent and the way their interactions are represented produces a new set of networks that goes beyond the concept of simple graphs, multilayer networks \cite{Kivela2014}.

\section{From simple graphs to multilayer networks}

The formalism of multilayer networks is an extension of the one for single networks (henceforth also called monoplex) i.e., networks that are completely described by just a set of entities (nodes) and their interactions (links). For this reason, we will begin this discussion providing a brief introduction to the notion of single graphs and then we will build the multilayer network framework from there. Nevertheless, we refer the interested reader to the books by Newman \cite{Newman2010} and Barabási \cite{Barabasi2016} for further details.

\subsection{Networks and graphs}

In its most general form networks are mathematically represented by graphs. A graph is a tuple $G=(V,E)$ where $V$ is the set of nodes (entities of the system) and $E \in V \times V$ is the set of links that connect pairs of nodes (two nodes are connected if they interact in some way). If $E$ is an unordered set the graph is said to be undirected whereas if it is an ordered set the graph is directed. In this last case, changes in the node where the link starts (source) will affect the node where it ends (target) but changes in the target node will not affect directly the source node. Note, however, that as we are treating the system as a whole, it may produce effects on the source node via other nodes. Additionally, in some cases, links have an associated number called weight which represents the intensity of the interaction.

Nonetheless, the usual representation of networks is the adjacency matrix. The adjacency matrix $\mathcal{A}$ of a monoplex graph is a $N\times N$ matrix (where $N$ is the number of nodes in the network) with elements $a_{ij}$ such that $a_{ij}=1$ if there is an edge between nodes $i$ and $j$ (or $w$ if the link has an associated weight) or $a_{ij}=0$ otherwise.

The most basic measure in networks is the degree of a node, defined as the number of links it possess,

\[k_i = \sum_{j=1}^N a_{ij}\]

We define the degree distribution of a network as the set of quantities $p_k$ which represent the fraction of nodes with degree $k$. It is possible to characterize networks by their degree distribution and it has been seen that almost all real-world networks have degree distributions with a tail of high-degree nodes.

The structural properties of the networks provide us a lot of information about their dynamics. For example, suppose that in the node $i$ there is an amount $\psi_i$ of some substance that is able to move along the links. The rate at which this substance flows from node $j$ to $i$ can be written as $C(\psi_j -\psi_i)$ where $C$ is a diffusion constant. Then, the rate at which $\psi_i$ is changing is given by
\[\frac{\mathrm{d} \psi_i}{\mathrm{d} t} = C\sum_j a_{ij} (\psi_j - \psi_i).\]

This equation can be rewritten as $\frac{\mathrm{d} \psi}{\mathrm{d} t} + C \mathcal{L} \psi = 0$. $\mathcal{L}$ is called the graph Laplacian and is defined as $\mathcal{L} = \mathcal{D} - \mathcal{A}$ where $\mathcal{D}$ is a diagonal matrix whose entry $i$ is $k_i$. The interesting aspect about this matrix is that its eigenvalues completely describe the diffusion process. 

Another important concept in network science is the notion of path. A path is a sequence of nodes such that every consecutive pair of them is connected by a link in the network. A particularly interesting example of path is the random walk. A random walk is a path where each step is chosen uniformly at random from the set of links attached to the current node. It can be shown that in the long time limit the probability of finding the random walk at node $i$ is proportional to its degree, $p_i = \frac{k_i}{2 m}$, where $m$ is the number of links in the network.

In the same way that the eigenvalues of the adjacency and laplacian matrices that describe the network can explain some of its dynamics, there are other measures of the structure of networks that are very useful to understand their behavior. One of the most interesting structural measures are those related to the concept of centrality. Centrality, in this context, refers to the importance of a node in the network, although the definition of importance can vary from one system to another. The degree itself can be regarded as a measure of centrality, known as degree centrality, as nodes with several links can produce effects in a large group of nodes. However, sometimes a node can be considered important not because of its intrinsic importance but because it is linked to other nodes that are themselves important. This is the concept behind eigenvector centrality, which can be obtained from the eigenvector corresponding to the largest eigenvalue of the adjacency matrix. Lastly, another interesting centrality measure is the betweenness centrality which is defined as the fraction of shortest paths between nodes $s$ and $t$ that pass through node $i$ out of the total number of shortest paths between those nodes. Thus, this quantity measures the extent to which a vertex lies on paths between other nodes.

Another interesting property of networks is that there are also intermediate scales that might play a key role for the functioning of the system at the global level. That is to say that not only individual nodes are important, but that it is also possible to find groups of nodes which share common properties. The simplest one is the group formed by three nodes. If all of them are connected to all others the relationship is said to be transitive. To quantify the level of transitivity in a network we define the clustering coefficient as the fraction of paths of length three in the network that are closed. Similarly, we might be able to classify nodes into groups or classes. If nodes of a given class tend to connect to nodes belonging to the same class we say that the network is assortative. It is usually useful to measure if the assortativity is larger than expected if nodes were randomly connected. Finally, another important structural property is related to whether or not so-called communities or modules can be identified in a network. A community or module is a set of nodes that are more densely connected among them than with nodes that are not in the same set (or module). There are several methods and heuristics for community detection, but the most popular ones are those based on maximizing the modularity of the network, which is defined as:

\[Q = \frac{1}{2m} \sum_{ij} \left(a_{ij} - \frac{k_i k_j}{2m} \right) \delta(c_i,c_j)\]
where $c_i$ is the class $i$ belongs to. 

Up to now, we have considered that the networks under study are isolated which is equivalent to assume that nodes are connected to each other by a single type of links and that other possible interactions not considered in the set $E$ are neglected. However, in many cases, this assumption can be an oversimplification. For example, neglecting time-dependence destroys the ordering of transmission processes. Similarly, ignoring the presence of multiple types of links might introduce errors when dynamical processes depend on the type of interaction. Thus, if we want to progress deeper in the study of complex systems we need to move to more advanced network structures, like multilayer networks.

\subsection{Multilayer networks}

To represent systems consisting of networks with multiple types of links, or other similar features, we consider structures that have layers in addition to nodes and links. In its more general form, in a multilayer framework a node $u$ in layer $\alpha$ can be connected to any node $v$ in any layer $\beta$. Layers will represent aspects or features that characterize the nodes or the links that belong to that layer. As a consequence, we can partition the set of links into  intra-layer links, that is links that connect nodes set in the same layer, and  inter-layer or coupling links which are those that connect nodes set in different layers.

Following de Domenico et al. \cite{de2013mathematical} we will use tensors, as a generalization of matrices, to represent these multilayer objects.  We can define the intra-layer adjacency tensor $W^\alpha_\beta (\tilde{k}) = \sum_{i,j=1}^N w_{ij}(\tilde{k}) E^\alpha_\beta (ij)$ as the one that indicates the relationships between nodes within the same layer $\tilde{k}$. Note that in unweighted monoplex networks $w_{ij}(\tilde{k}) = a_{ij}$ which are the entries of the adjacency matrix. From now on we will use the tilde symbol to differentiate indices that correspond to nodes from those that correspond to layers.

\begin{figure}
\includegraphics[width=\linewidth]{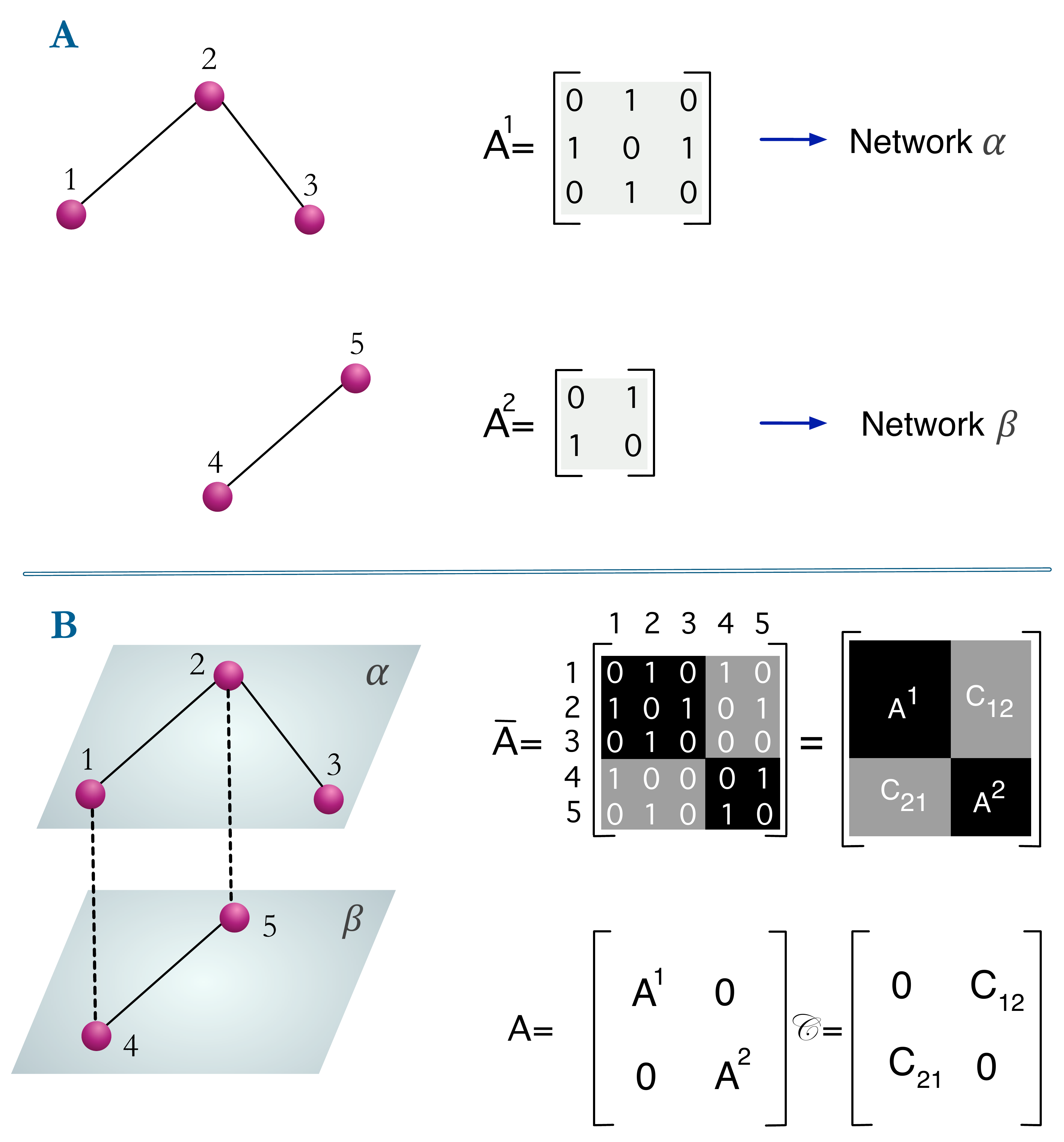}
\caption{The figure shows monoplex networks (panel A) and a multilayer system (panel B) made up of the two monoplex networks i (A). Panel A: the interactions between the first set of nodes (1,2,3) are not affected and do not affect the second set of nodes (4,5). Therefore, each system corresponds to a single layer network, whose adjacency matrices are shown. Panel B: When the sets of nodes are not independent and they affect each other, a multilayer representation is more accurate. In the example shown, the structure of each layer is represented by an adjacency matrix  $\mathcal{A}^{(i)}$, where $i=\{\alpha=1, \beta=2\}$. $\mathcal{C}_{(\alpha\beta)}$ stores the connections between layers $\alpha$ and $\beta$. Note that the number of nodes in each layer is not the same.}
\label{fig:1}
\end{figure}

To incorporate the possibility of a node in one layer $\tilde{h}$ to be connected to another one in layer $\tilde{k}$, we introduce the inter-layer adjacency tensor $C^\alpha_\beta (\tilde{h} \tilde{k})$, which corresponds to $W^\alpha_\beta (\tilde{k})$ if $\tilde{h} = \tilde{k}$. Lastly, by denoting the canonical basis as $E^{\tilde{\gamma}}_{\tilde{\delta}} (\tilde{h} \tilde{k}) = e^{\tilde{\gamma}}(\tilde{h}) e_{\tilde{\delta}} (\tilde{k})$ we can write the multilayer adjacency tensor as
\[M^{\alpha \tilde{\gamma}}_{\beta \tilde{\delta}} = \sum_{\tilde{h}, \tilde{k} = 1}^L C^\alpha_\beta (\tilde{h} \tilde{k}) E^{\tilde{\gamma}}_{\tilde{\delta}} (\tilde{h} \tilde{k})\]
\noindent
where $L$ is the number of layers. Alternatively, it is possible to extend the adjacency-matrix representation to encode multilayer networks making it easier to use tools that have already been developed to study monoplex networks also in this new framework. We can do so by building a block matrix where each diagonal block corresponds to the adjacency matrix of each layer, $A^\alpha$, and the off-diagonal matrices encoding the coupling, $C^{\alpha \beta}$, see figure \ref{fig:1}.

Multilayer networks encode two major classes of systems \cite{bianconi2015interdisciplinary}: multiplex networks and network of networks. Multiplex networks are networks where the same set of nodes is represented in every layer, although the interaction between nodes might be different in each one. As an example, two nodes might be connected in one layer an might not in other. This is the case of online social systems, where a given user might have a Twitter account (layer 1) and a Facebook profile (layer 2). The set of followers/friends does not in general coincide for both layers, thus leading to two different intra-layer adjacency matrices. On the contrary, a network of networks is instead formed by networks that are interlaced to each other but formed by different types of nodes. Although these two are the most common kinds of multilayer networks, Kivel\"{a} et al.\cite{Kivela2014} present several other types and denominations of multilayer networks.

When building multilayer networks it is often not completely clear how to define each layer, the interactions among them or even how many of them are necessary. De Domenico et al. \cite{de2015structural} tackle the problem of reducibility, i.e., defining the number of layers a multilayer network needs to have to accurately represent the structure of the system. Using the Von Neumann entropy on a multilayer network it is possible to determine if the multilayer representation is distinguishable from the aggregated network. This way, they propose that if the aggregation of two layers does not result in a decrease of the relative entropy with respect to the multiplex where they are separated, they should be kept aggregated. Likewise, Menichetti et al. \cite{menichetti2014weighted}, propose a measure of the amount of additional information that can be extracted from multiplex networks over the one contained in the individual layers separately based on the entropy of multiplex ensembles. Morover, Kleineberg et al. \cite{kleineberg2016hidden} show that multiplex networks are not just random combinations of single network layers but rather posses significant hidden geometric correlations. Finally, Cozzo et al. \cite{cozzo2016multilayer} showed that as a function of the coupling between layers, different multiplexity regimes can be identified from the spectral properties of the graph.

The extension of the tools already developed for single layer networks to the multilayer framework is sometimes straightforward \cite{boccaletti2014structure}. For example, we can define the degree of node $i$ in layer $\alpha$ as $k_i^\alpha = \sum_j a_{ij}^\alpha$. Consequently, its degree in the multilayer network is no longer a scalar but the vector $k_i = \{k_i^1,\ldots,k_i^L\}$, which results in a total degree or degree overlap of $o_i = \sum_\alpha k_i^\alpha$. It is also possible to create new measures like the edge overlap, $o_{ij} = \sum_\alpha a_{ij}^\alpha$, which accounts for the number of layers where the same link exists. Besides, as nodes are now characterized by  vectors instead of scalars, we need to develop new tools to simplify their description. For instance, we can quantify the distribution of the degree of a node among the various layers using the entropy of the multiplex degree,
\[H_i = - \sum_{\alpha=1}^L \frac{k_i^\alpha}{o_i} \ln\left(\frac{k_i^\alpha}{o_i}\right)\]

\noindent
or the participation coefficient \cite{battiston2014structural}.

There are other measures similar to the edge overlap that only work under the multilayer framework, for instance the interdependence \cite{battiston2014structural}. The interdependence of node $i$ is defined as $\lambda_i = \sum_{i\neq j} \frac{\psi_{ij}}{\sigma_{ij}}$ where $\sigma_{ij}$ is the total number of shortest paths between nodes $i$ and $j$ and $\psi_{ij}$ is the number of shortest paths between node $i$ and node $j$ that makes use of links in two or more layers. Thus, it measures how dependent is a node on the multiplex structure in terms of reachability. Equivalently, it is possible to extend this definition from nodes to layers to account for the importance of a given layer in the whole system \cite{aleta2017multilayer}.

Cozzo et al. \cite{cozzo2015structure} extended the notion of triadic relations, the transitive relationship commented previously, to multiplex networks. In particular, they distinguish 5 different triadic relations: one where the three links are in the same layer, three where two links are in one layer and the third is in a different layer and finally the one where the three links are in three different layers. 

Similarly, to study assortativity in multilayer networks there are several possibilities: we can analyse the assortativity between nodes in different layers, between layers themselves or inside each layer separately, etc. De Arruda et al. \cite{de2016degree} showed that if the multilayer adjacency tensor is expressed as $M^{\alpha \tilde{\gamma}}_{\beta \tilde{\delta}} = \sum_{\tilde{h},\tilde{r}}^L C^\alpha_\beta (\tilde{h}\tilde{r}) E^{\tilde{\delta}}_{\tilde{\gamma}} ( \tilde{h}\tilde{r})$, where $C^\alpha_\beta (\tilde{r}\tilde{r})$ is the adjacency matrix for layer $\tilde{r}$, the assortativity coefficient can be written as
\[\rho(\mathcal{W}^\alpha_\beta)=\frac{\mathcal{M}^{-1} \mathcal{W}^\alpha_\beta Q^\beta Q_\alpha - [0.5 \mathcal{M}^{-1} (\mathcal{W}^\alpha_\beta Q_\alpha u^\beta + \mathcal{W}^\alpha_\beta Q^\beta u_\alpha)]^2}{\mathcal{M}^{-1} (\mathcal{W}^\alpha_\beta (Q_\alpha)^2 u^\beta + \mathcal{W}^\alpha_\beta (Q^\beta)^2 u_\alpha) - [0.5 \mathcal{M}^{-1} (\mathcal{W}^\alpha_\beta Q_\alpha u^\beta + \mathcal{W}^\alpha_\beta Q^\beta u_\alpha)]^2}.\]

Thus, by varying the definition of $\mathcal{W}^\alpha_\beta$ it is possible to study the assortativity in each layer, with $\mathcal{W}^\alpha_\beta = C^\alpha_\beta (\tilde{r} \tilde{r})$, the assortativity of the projected network, with $\mathcal{W}^\alpha_\beta = M^{\alpha \tilde{\delta}}_{\beta \tilde{\gamma}} U^{\tilde{\gamma}}_{\tilde{\delta}}$, the assortativity of selected sets of layers, with $\mathcal{W}^\alpha_\beta (\mathcal{L}) = M^{\alpha \tilde{\delta}}_{\beta \tilde{\gamma}} \Omega^{\tilde{\gamma}}_{\tilde{\delta}} (\mathcal{L})$ where $\Omega^{\tilde{\gamma}}_{\tilde{\delta}} \mathcal{L}$ is the unity tensor when $\tilde{\gamma}$ and $\tilde{\delta}$ are selected and zero otherwise, or the correlation between different layers with $\mathcal{W}^{\tilde{\gamma}}_{\tilde{\delta}} = M^{\alpha \tilde{\gamma}}_{\beta \tilde{\delta}} U^\beta_\alpha$. We also note that there are several other types of correlations that can be analysed in complex networks, see for example Nicosia et al \cite{nicosia2015measuring}.

Following our previous discussion, random walks are often used to explore a network using only local information due to their simplicity. Particularly interesting are biased random walkers whose transition probabilities depend on the topological properties of the destination node, as they can be used to define centrality measures, identify communities or provide optimal exploration of a network. However, it is not straightforward to extend the notion of random walks from simple networks to multilayer networks. Battiston et al. \cite{battiston2016efficient} explore the consequences of redefining random walks in multiplex networks. Note that the richness of multilayer networks in this context comes from the fact that node properties are no longer scalar but vectorial, like their degree, which allows to define more complex biasing functions.

Likewise, random walks can be used to define centrality measures in multilayer networks \cite{sole2016random}. Other centrality measures used in single layer networks can also be extended to the framework of multilayer networks like PageRank \cite{tu2018novel}, betweenness \cite{sole2014centrality} or eigenvector centrality \cite{sola2013eigenvector, Buldu2016}. As de Domenico et al. \cite{de2015ranking} showed, when one computes centrality measures in the context of multilayer networks, it is possible to find versatile nodes which cannot be extracted directly from the aggregated network. Furthermore, it is possible to use dynamical processes to characterize the relative position of nodes in multiplex networks \cite{reiffers2017opinion}.

As briefly discussed previously, the spectral properties of networks play a very important role in the dynamics they represent. Cozzo et al. \cite{Cozzo2013Nov}, using perturbation theory, showed that the largest eigenvalue of the multiplex network is  equal to the one of the adjacency matrix of the dominant layer of the system at a first order approximation. Similarly, S\'{a}nchez-Garc\'{i}a et al. \cite{sanchez2014dimensionality} studied thee eigenvalues of the Laplacian in the context of multilayer networks. They show that the eigenvalues of the quotient (a coarsening of the original network) are interlaced with the eigenvalues of its parent network. This fact has deep consequences as, for example, the relaxation time on a multiplex network is at most the one of the aggregated network, which in turn can result in faster diffusion processes on multiplex networks than in their aggregated counterparts.

Finally, we will examine the state of community detection in multilayer networks. Recall that in some networks nodes tend to cluster into groups giving rise to non-trivial structures. Mucha et al. \cite{mucha2010community} generalized the concept of modularity considered previously to multilayer networks, which can be used to detect communities not only in static networks but also in time-dependent networks by setting the state of the network at each time window in one layer. There are other extensions of modularity \cite{Pramanik2017Oct} as well as other methods of community detection like flow based processes \cite{de2015identifying} or random walks \cite{jeub2017local, kuncheva2015community}. Methods have been developed also to detect overlapping communities \cite{wilson2017community} and study the detectability \cite{Taylor2016Jun} of community structures in stochastic block models \cite{Bacco2017Apr, Peixoto2015Oct, Valles-Catala2016Mar}.

Next, we also summarize the main results obtained so far for several paradigmatic dynamical processes when they take place on top of multilayer networks.

\section{Dynamical processes on multilayer networks}

This section is divided in two subsections. First, we will explore the problem of percolation which is one the first phenomena that were tackled by condensed matter physicists. Even though its study started in the 1940s and there is a large theoretical framework already built, it is still very relevant today due to its importance in the description of networks. Indeed, not only it can be used to analyze the resilience of networks, which would be the most straightforward application of the theory, but it can also help study problems found in very different fields, being vaccination and herd immunity two relevant applications on health sciences \cite{Newman2010}. Then, we will round off this section with the description of diffusion processes in multilayer networks.  We will also provide some information about problems where diffusion processes on networks are quite relevant, like the cases of epidemic and opinion spreading.

\subsection{Percolation and multilayer networks}

In undirected single layer networks, a connected component is the maximal set of nodes that are all connected to one another via some path. In multilayer networks, this definition can be trivially extended by including links that connect layers, so that the set of nodes is connected to each other within each layer and also connected to the rest of the layers \cite{lee2015towards}. Even though the extension is simple, by incorporating more layers to the system the number of nodes can grow quite fast. Thus, it is important to implement new algorithms for the detection of connected components in multilayer networks \cite{schneider2013algorithm, hwang2015efficient}.

One of the main applications of percolation theory in networks is the study of robustness. Robustness in this context refers to the ability of preserving the structure of the network when it is subject to failures or attacks, either in the nodes or in the links. Although the robustness of single layer networks has been deeply analyzed and the main questions are already answered, in the case of multilayer networks there are new and exciting challenges. 

For example, in contrast to single layer networks where under random removal of nodes the size of the largest connected component has a continuous transition, Baxter et al. \cite{baxter2012avalanche} showed that in multiplex networks the transition is hybrid: below the critical point there is a discontinuity like a first-order transition, but above it, the systems exhibits critical behavior typical of a second order transition. Similarly, another property that behaves in a different way when studying robustness in single layer networks and multilayer networks is clustering \cite{Danziger2016}. In single layer networks the effect of clustering in the robustness of the network is very low. However, Shao et al. showed that clustering substantially reduces the robustness of some multilayer networks \cite{Huang2013Jan, Shao2014Mar}.

Buldyrev et al. \cite{buldyrev2010catastrophic} studied an interdependent network of power grids and computers. In this context, we can regard inter-layer links as dependency links. That is, for a node to function in one layer, it requires support from another node that is in a different layer so that it depends on the inter-layer link. In this particular case, the shutdown of power stations led to the failure of nodes in the communication network which in turn caused further breakdown of power stations. They modeled this process with an interdependent network with two layers and showed that these networks can be less robust than single networks. Indeed, a very important node in one layer might be connected to a smaller, more fragile, node in the other one. If this last node disappears, it will radically affect the first layer as it will shutdown the node that it is important. 

There are multiple extensions for these models of failures in interdependent networks. For instance,  Min et al. \cite{min2014multiple} addressed the problem when the failure of a node in one layer does not directly cause a failure in a node in the other one, but instead depletes the resources that the node gets. If that value gets below a certain threshold then the node fails. With this model they found that the system exhibits hysteresis which would produce an increase in the cost of the recovery process. In a similar way, Shao et al. \cite{Shao2011Mar} built networks where nodes are connected by more than one link to other layers so that the failure of just one node does not always produce breakdowns in the other layer. Conversely, Son et al. \cite{son2012percolation} use an epidemic spreading approach to address the problem of percolation in interdependent uncorrelated networks.

These processes in interdependent networks are also known as cascading processes. The failure of a node in layer $A$ will shutdown some nodes in layer $B$. Consequently, this will produce even more failures in layer $A$. This process will continue iteratively until all affected nodes are removed from the network. We also note here that from a numerical point of view, the algorithms used for these models can be sometimes too slow, but there are new algorithms proposed to handle these kind of problems specifically in multilayer networks \cite{grassberger2015percolation}.

Not only interdependent networks have been addressed. Other authors like Zhao et al. \cite{zhao2016robustness} have explored the robustness of multiplex networks when their nodes are removed, either by random failures or by targeted attacks. Besides, it is also very important to determine the effects that each network property has on the robustness of networks. For instance, Cellai et al. \cite{cellai2016message} proposed a percolation model to handle multiplex networks with link overlap based on message passing . Min et al. \cite{min2014network}  analyzed the problem of network robustness when there are interlayer degree correlations and Lee et al. \cite{lee2016strength} addressed the problem from the point of view of layers in the context of the international trade network.

Finally, we conclude this section by giving some examples of applications of these percolation processes to networks coming from very different fields. Reis et al. \cite{Reis2014Sep} developed a theory to study stability of interconnected networks and demonstrated it in the context of brain networks. Brummitt et al. \cite{brummitt2015cascades} studied bankruptcy spreading using cascades in multiplex financial networks. Lastly, Baggio et al. \cite{Baggio2016Nov} quantified the robustness of socio-ecological systems under plausible scenarios of change and found that changes in social relations precipitated steeper declines in network interconnectedness than changes in ecological resources.

\subsection{Diffusion processes in multilayer networks}

Diffusion processes have long been studied in the context of networks. It has been shown that stochastic processes such as evolutionary games or epidemic spreading present very different behavior when applied on populations structured in networks than when there is no underlying structure. For instance, the epidemic threshold, which is a key quantity in mathematical epidemiology, vanishes under some network configurations. Thus, it is a must to know the underlying structure of the population where the process is taking place. Besides, as the structure of multilayer networks encodes new structural interdependencies, it is necessary to study what new effects they induce in the dynamical behavior of the system.

The first problem we need to address is how to define dynamics on these networks. First, we can have the same dynamics in every layer so that the only difference is the connection pattern found in each layer. As we stated previously, the main characteristics of the diffusion process are controlled by the Laplacian tensor. In particular, the timescale is controlled by the second smallest positive eigenvalue of the supra-Laplacian matrix (which is obtained by flattening the Laplacian tensor) . Interestingly, for some parameters its value is such that the diffusion process is faster in the multiplex network than in the separated layers \cite{Domenico2016Aug}. Conversely, it is possible to implement a different dynamical process in each layer so that they can enhance or inhibit each other as we shall see later.

The two main diffusion processes that are studied in multilayer networks are epidemic spreading and information diffusion. In fact, sometimes the same equations are used for both processes, as in some cases they are very similar. As an example of opinion dynamics,  we mention the study by Amato et al. \cite{Amato2017Dec}, in which the authors studied the competition of opinions in a two layer network by allowing the same individual to have a different opinion in each layer, representing in this way the possibility of individuals having different opinions depending on the context (layers represent diverse scenarios). 

As noted before, epidemic-like processes have also been used to study information dissemination. We thus focus on models for disease spreading on the remaining of this section. The addition of multiple layers to the system can have dramatic effects on the outcome of an epidemic. For example, Buono et al. \cite{Buono2014Mar} showed that with partial overlap, i.e., with only a fraction of the nodes in one layer being also in the other one, both the epidemic threshold and the total fraction of infected individuals significantly change even for low values of overlap.

To show in a straightforward way the differences between studying epidemic processes on a single layer network and on a multilayer network we will use the supra-adjacency matrix representation of this last system. Let us suppose that we want to study a susceptible-infected-susceptible (SIS) model in a simple network. That is, nodes can be susceptible (S) and get infected (I) with probability $\beta$ when they contact infected individuals. Once a node becomes infected, it might recover with probability $\mu$ and get into de susceptible compartment again. We can express this process using a Markov chain approach as \cite{gomez2010discrete}
\[p_i(t+1) = (1-p_i(t))(1-q_i(t)) + (1-\mu) p(t), \]

\noindent
where $p_i(t)$ is the probability of node $i$ being infected at time $t$ and $q(t)$ is the probability of node $i$ not getting infected by any neighbor, which is given by 
\[q_i(t) = \prod_{j} [1-\beta R_{ij} p_j(t)],\]

\noindent
in this context, $R_{ij}$ is the contact probability matrix,
\[R_{ij} = 1-\left(1-\frac{a_{ij}}{k_i}\right)^\lambda.\]

\noindent
With this formulation one can go from a contact process (one contact per unit time) when $\lambda=1$ to a fully reactive process (all neighbors are contacted) when $\lambda \rightarrow \infty$. Note that $a_{ij}$ stands for the element $ij$ of the adjacency matrix. Thus, it is in this last element where the structure of the network plays a role.

This equation hast always the trivial solution $p_i=0 \forall i$. Other nontrivial solutions can be easily computed numerically by iteration. Nevertheless, it is possible to get a condition to determine if it is possible to have nontrivial solutions or not. Indeed, linearizing $q_i$ around 0 we get
\[\left[R-\frac{\mu}{\beta}I\right]p = 0\]

\noindent
which has nontrivial solutions if and only if $\frac{\mu}{\beta}$ is an eigenvalue of $R$. Thus, the epidemic threshold (i.e. the critical point) will be
\[\left(\frac{\beta}{\mu}\right)_c = \frac{1}{\Lambda_{\mathrm{max}}}\]

\noindent
where $\Lambda_{\mathrm{max}}$ is the largest eigenvalue of the matrix $R$.

Now, following Cozzo et al. \cite{Cozzo2013Nov} if we have a multiplex system we can define the supracontact probability matrix as
\[\bar{R} = \bigoplus_\alpha R_\alpha + \left(\frac{\gamma}{\beta}\right)^T C\]

\noindent
where the $R_\alpha$'s are the contact probability matrices of each layer $\alpha$ and $C$ is the interlayer coupling matrix. Note that we have to divide by $\beta$ the second term of the equation because if we did not, once we multiply $\bar{R}$ term by $\beta$ we would be spreading the disease in the same way both in intra-layer links and inter-layer links. That is, we would be losing the distinction between links inside layers and links that connect nodes of different layers. Likewise, $\gamma$, which represents the probability of transmitting the disease through inter-layer nodes, needs to be also inside the definition of the supracontact probability matrix.

The power of this formulation is that we can use the exact same equations as in the previous case just by changing the contact probability matrix $R$ by the supracontact probability matrix $\bar{R}$. Hence, the epidemic threshold in this framework is given by
\[\left(\frac{\beta}{\mu}\right)_c = \frac{1}{\bar{\Lambda}_\mathrm{max}}\]

At first glance, it may seem that nothing has changed, but there is an important difference. Indeed, the largest eigenvalue value is no longer fixed by the network structure, but depends on the ratio $\gamma/\beta$. Even more, it is worth analyzing this result by means of a perturbative approach by fixing the ratio $\epsilon = \gamma /\beta \ll 1$. By doing so, at first order approximation the largest eigenvalue of $\bar{R}$ is $\bar{\Lambda}_\mathrm{max} = \mathrm{max}_\alpha \{\Lambda_\alpha \}$ and hence the emergence of a macroscopic steady state is determined by the layer with the largest eigenvalue. This is the dominant layer as it sets the critical properties of the entire multilayer system.

\begin{figure}
\includegraphics[width=\linewidth]{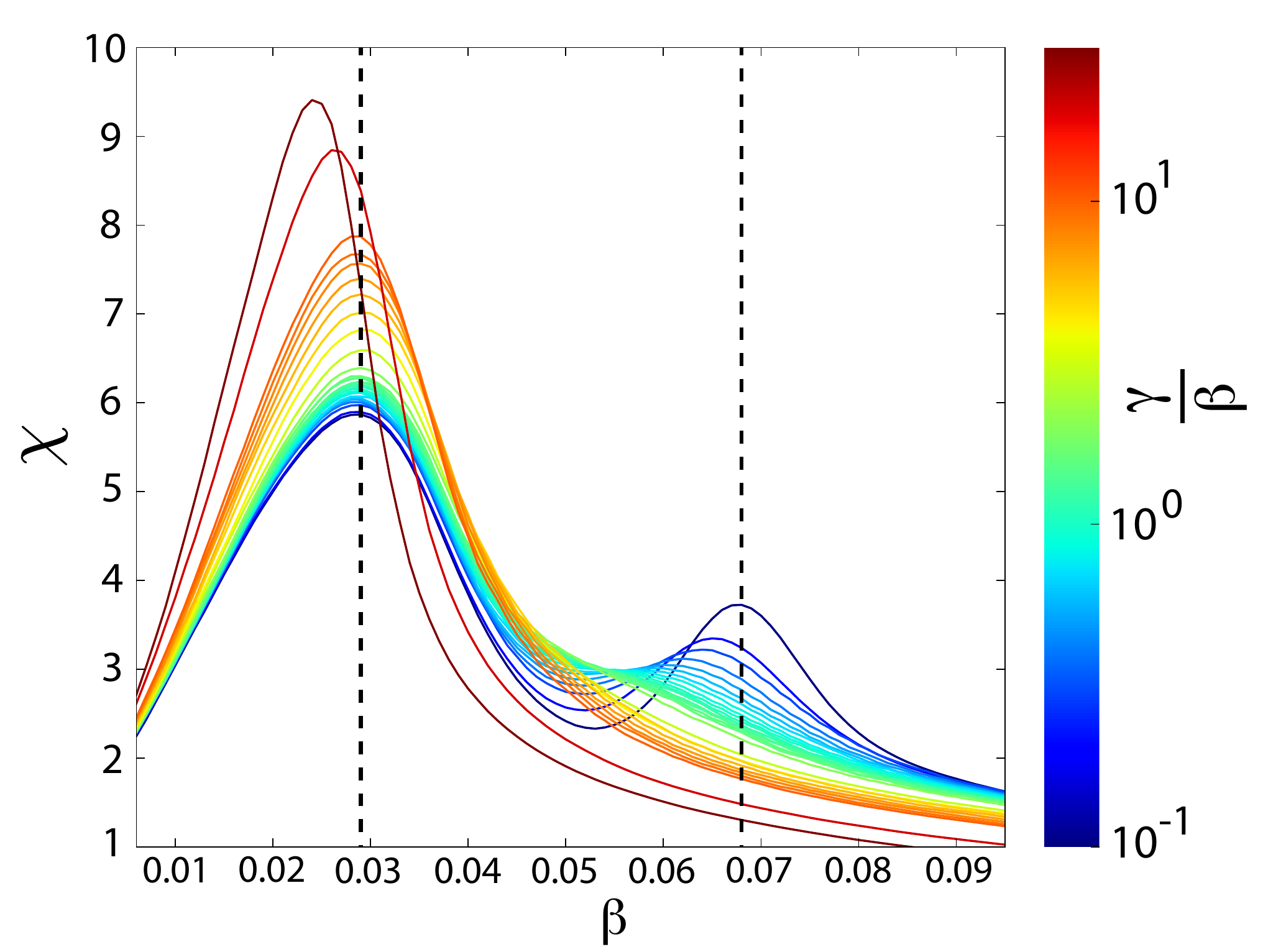}
\caption{The figure shows the value of the susceptibility $\chi$ as a function of the probability of infection $\beta$ and the value of the coupling between layers $\gamma/\beta$. There is a peak of the susceptibility at the phase transition of the system, i.e., when the system goes from a disease free state to an endemic one. Interestingly, for a certain set of parameters there are two phase transitions instead of one. Indeed, for low values of the infection probability and the coupling it is possible that the disease is able to spread through the whole network but unable to reach other layers (first transition). If the infection probability is increased, the disease will be able to transition from said layer-localized state to a delocalized state (second transition). This figure has been adapted from \cite{Arruda2017Feb} with the permission of the authors.}
\label{fig:2}
\end{figure}

Furthermore, de Arruda et al. \cite{Arruda2017Feb} studied the whole spectrum of this object and found that there are multiple transitions. Specifically, for a 3-layer network, the first transition changes the system from a disease free state to an endemic one but layer-localized, i.e., the epidemic is confiend to only one layer. The second point represents the transition from a layer-localized state to a delocalized state, see figure \ref{fig:2}. We also mention that a similar approach was used by Valdano et al. \cite{Valdano2015Apr} to develop a theoretical model to investigate epidemic processes in arbitrary temporal networks.

As we have seen the introduction of inter-layer links is not straightforward and needs to be done with care as they can represent several things. For instance, Min et al. \cite{Min2016Feb} studied a model of information spreading through multiple interaction channels (layers) subject to layer-switching costs. They demonstrated that these costs can affect the outcome in non-trivial ways so that results cannot be reproduced by using a single aggregated layer. Conversely, Alvarez et al. \cite{Zuzek2015Sep} set the transmission probability on inter-layer links to 1 so that they only play the role of connecting the layers (there is no actual transmission through them).

As noted before, these kind of models are not only interesting on their own. Indeed, multilayer networks allow to study a plethora of more complex processes that can be naturally represented as multilayer systems. For instance, \cite{Kouvaris2015Jun} investigated pattern formation induced by diffusive transport in multiplex networks. Similarly, Biondo et al. \cite{Biondo2017Nov} built a network composed by a trading layer and an information layer. In this way, the trading patterns of individuals, which are encoded in the first layer, are affected by the spreading of information produced in the second one. Lastly, Nicosia et al. \cite{Nicosia2017Mar} built a multilayer network mimicking neural dynamics in one layer and nutrient transport in the other one. In their model, the diffusion of nutrients in one layer affects the level of synchronization in the other one and is even able to produce explosive synchronization that layer. 

It is clear then that diffusion processes can be used to study a large variety of systems, which explains their popularity. In the last section we will explore some of these systems as we will briefly describe how the framework of multilayer networks is being used in several fields of knowledge.

\section{Other examples of multilayer networks}

Here we provide a brief account of some applications in different areas where the multilayer methodology has been used recently. We will not cover these subjects deeply, but instead we provide references to the most recent reviews on each subject so that the interested reader can easily find its way.

\subsection{Multilayer networks in ecology}

Natural systems typically exhibit multiple types of interactions, such as the same plant species interacting with both pollinators and herbivores. Thus, much can be gained by disentangling these interactions and organizing them in multilayer networks. Indeed, as Pilosof et al \cite{Pilosof2017Mar} state, there are some multilayer layerings that can provide deep insight into the behavior and evolution of ecological systems. In particular, layers can represent the same system at a different point of space or time, which will be useful when studying long term evolution of ecological systems. If we center our attention on links, we can define each layer according to the interaction type. Conversely, if we are more interested in the organization of nodes, we can define layers by dome group identity or by the level of organization within the system.

For instance, Finn et al. \cite{finn2017} used the third approach to examine the role of individuals in a society of baboons. In this way, they built a multilayer network of two layers: one represents grooming while the other represents association (i.e. proximity). Then, they studied the centrality of the baboons using this representation and compared it to the results obtained for the aggregated (single layer) network. Interestingly, they found that the most central individual in the multilayer representation is not that central in the aggregated one and it is not central at all in the individual layers. They concluded that when studying relationships that depend on multiple interaction types, it might be necessary to use the multilayer framework to capture individual's social roles.

\subsection{Multilayer networks in biology}

In the recent review by Gosak et al. \cite{Gosak2018Mar} it is summed up the state of network science in the study of biological systems at different scales. In addition to giving an overview of the use of single layer networks in biological systems, they also discuss the possibilities the multilayer network formalism provides. For instance, one can build a protein-protein interaction multilayer network where each layer represents the life stage of a bacteria \cite{Shinde2015Dec}. But this research is no longer constrained to small and simple systems. As an example, Zitnik et al. \cite{Zitnik2017Jul} built a multilayer network of molecular interactions where each layer represents a different human tissue. By imposing a tissue hierarchy, they were able to improve the predictive power of this description compared to the single layer case.

\subsection{Multilayer networks in transport}

Transportation systems are one particular example of those systems where the multilayer formulation arises in a natural way as there can be several modes of transport between two given locations. These modes can have very different properties such as velocity or carrying capacity and thus it is important to be able to distinguish each of them when studying the whole system. 

\begin{figure}
\includegraphics[width=\linewidth]{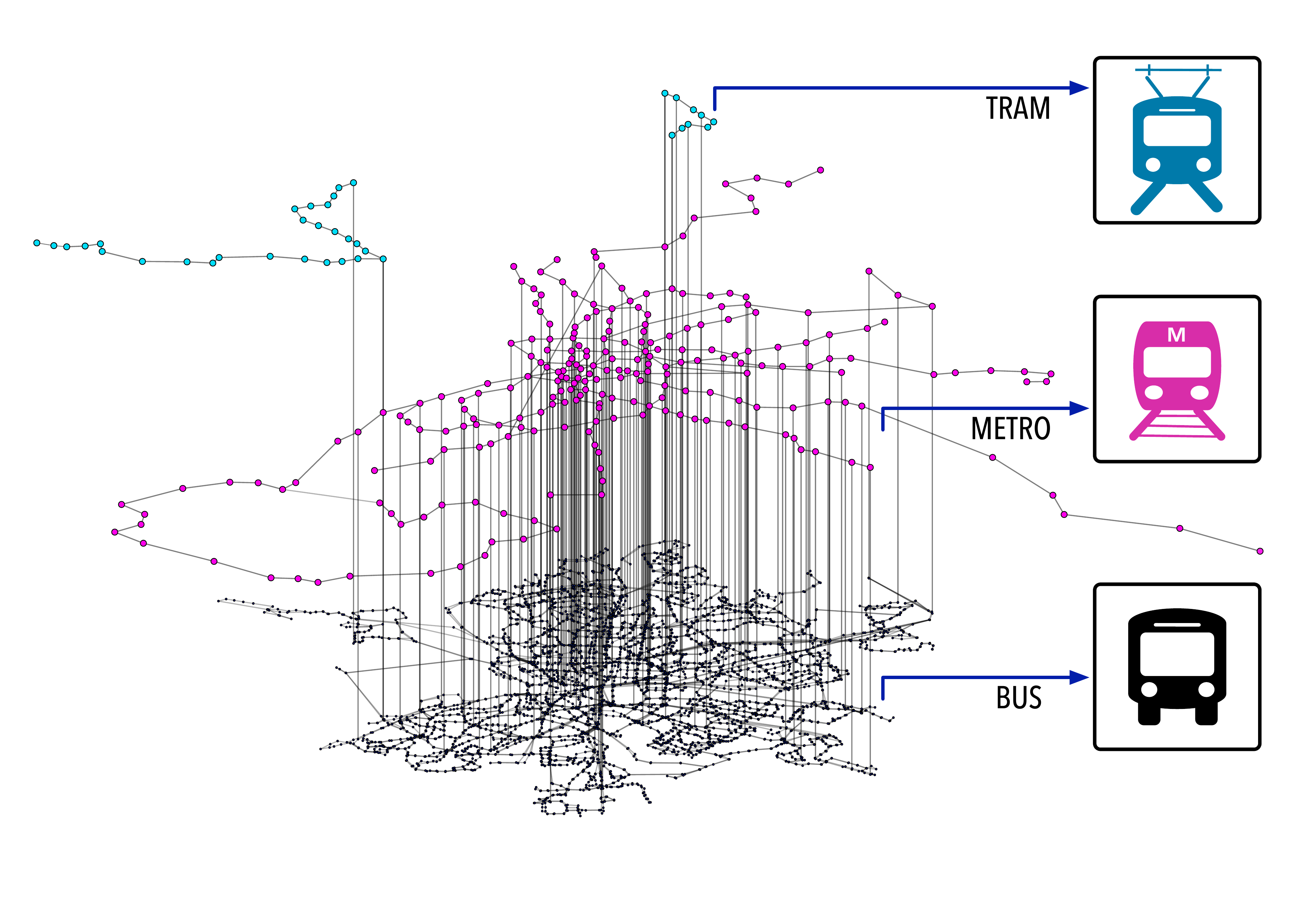}
\caption{The figure shows the multilayer representation of the transport system of Madrid. The first layer (blue nodes) represents the tram system, the second layer (green nodes) the metro system and the third layer (red nodes) the bus system. Vertical links connect stops of different transport modes that are within a 150m radius. This figure has been adapted from \cite{aleta2017multilayer} with permission of the authors.}
\label{fig:3}
\end{figure}

There are several ways of modeling transport systems as multilayer networks depending on the aim of the study. For instance, at a country level one can model each transport mode, i.e., coach, trains, flights, etc. as layers and cities as nodes \cite{gallotti2015multilayer}. A similar approach can be used when studying smaller systems like urban transport, where nodes might be locations in the city and each layer might represent a mode of transport such as buses, metro, rail, etc. \cite{gallotti2014anatomy}, see figure \ref{fig:3}. However, in this last type of systems it is important to not only analyze the interplay between different modes but also between lines of each mode. Indeed, to properly take into account the transfer time from, for instance, one metro line to another it is necessary to consider each line as a single layer. Then, layers of each transport mode can be grouped into a higher level entity, superlayers, depending on the properties one is interested to unveil \cite{aleta2017multilayer}. Besides, as urban transports are embedded in the city, it might also be interesting to add the street pattern as a layer in order to provide novel insights into urban planning \cite{strano2015multiplex}.

Similarly, the multilayer methodology has also been applied to study flight networks. In particular, Cardillo et al. \cite{cardillo2013modeling} built the european air transport multilayer network where each layer represented an airline. In this way, they showed that the resilience of the multilayer network is lower than the one of the aggregated network. Other air transportation networks that have been analyzed using the multilayer framework include the greek aviation network \cite{tsiotas2015decomposing} and the chinese aviation network \cite{hong2016structural, jiang2016network}.

\subsection{Multilayer networks and the human brain}

The traditional tools used to analyze the architecture of the human brain have been focused on  single scales. Following Betzel et al. \cite{Betzel2017Oct}, we can distinguish three scales in brain networks, namely: the spatial scale, which refers to the granularity at which the network is defined; the temporal scale, ranging from sub-miliseconds to the entire lifespan; and the topological scale, ranging from individual nodes to the network as a whole. Most analysis of the human brain network done so far fix these three characteristics to a given value. However, now it is possible to address problems like the time-varying connectivity of the brain using multilayer networks. Furthermore, if one tackles the same problems, as identifying central nodes, in single layer brain networks and multilayer networks the results are quite different \cite{de2017multilayer}.

\subsection{Multilayer networks in economy}

The multilayer formulation allows for novel approaches to the study of the dynamics of the economy. Musmeci et al. \cite{musmeci2017multiplex} applied the multilayer approach to study the structure of financial markets by building a multilayer network where each layer represents the same data but the links are constructed using different correlation measures: Pearson, Kendall, Tail and Partial correlation. In this way they can provide a complete picture of the market dependency structure as the Pearson layer accounts for linear dependencies, the Kendall layer for monotonic nonlinearity, the Tail gives information about correlations in the tails of the distributions and finally the Partial correlation detects direct relationship that are not explained by the market.

There are several other examples of the multilayer framework being applied to study economical systems. Bargigli et al. used it to study the interbank network \cite{bargigli2015multiplex}, Zeng et al. performed a multilayer network analysis of EU lobby organizations \cite{zeng2016multiplex}, Condorelli et al. studied dynamical model of bilateral trading in multilayer networks \cite{condorelli2016bilateral}, Santana et al. studied investor commitment to startups and entrepreneurs \cite{santana2017investor} and Battiston et al. described financial systems as interconnected networks \cite{battiston2016financial}.

\subsection{Multilayer networks in game theory}

As multilayer networks can account for several different social contexts at the same time, they are becoming quite popular in the understanding of human cooperation \cite{wang2015evolutionary}. A prominent example is interdependent network reciprocity, which is capable of maintaining healthy levels of cooperation even under extremely adverse conditions \cite{wang2013optimal}. However, as Battiston et al. state \cite{battiston2017determinants}, this particular field is still in its infancy and it is important not to be overly optimistic with these results. In particular, they showed that network interdependence can effectively promote cooperation past the limits imposed by isolated networks, but this only occurs if the multilayer network fulfills some specific conditions. Thus, the involvement in different social contexts on its own is not sufficient for the promotion of cooperation.

\section{Outlook}

In this review, we have provided a brief and informal introduction to the subject of multilayer networks. Most of the works in this field have been done since 2012, and yet an exhaustive review will likely produce several hundreds pages. This illustrates the potential of the new methodology to tackle real world problems. Here we have introduced the reader to a few of such problems, but the list of possible applications is still growing. Among the open challenges and promising venues for the use of the multilayer formalism, we would like to mention just a few as a way to round off this partial revision.

First, there are still some fundamental problems to address regarding the theoretical description of multilayer networks. The vast majority of works done up to date have dealt with the simplest network kinds one can consider, that is, undirected and unweighted networks in each layer. While several results can be shown to hold for the case of weighted multilayer networks, this is not so when one deals with directed networks. In this latter scenario, the adjacency or supra-adjacency matrices are not symmetric, which makes their study harder, for instance, in terms of the spectral properties of the system. Secondly, there is still much to do regarding adaptive and temporal multilayer networks. These latter cases could describe systems as relevant as the brain, social networks in which new connections are created and old ones are deleted as time goes on, or even when one studies dynamics in which nodes can react and adapt in many different ways (which, for instance, might need to work with diffusion coefficients that are time dependent). 

As for new applications, the use of multilayer networks to study complex biological systems is a promising direction. At the cellular level, proteins and genes are often involved in many signaling, transcriptional or metabolic regulations -which in turn depend on external stimulus or the environmental conditions- that take place concurrently, which calls for the need to use multiple channels to describe each of these pathways, that is, a multilayer framework. At the population level, a field that could greatly benefit from the new formalism is that of disease spreading. This is the case of an increasing number of real scenarios in which two or more diseases propagate on top of the same host population, giving rise to complex dynamical interdependencies like cross-immunization or cooperation between the diseases $-$as it is the case of Tuberculosis and HIV$-$ that can be accounted for considering multiple networks of contacts.

In summary, despite the many advances of the last years in the field of multilayer networks, we expect that the field continues to grow, both in terms of foundational research as well as in their application to several problems in many different areas of research, from the social and biological sciences to natural and technological systems.

\newpage

\bibliography{referencias}
\bibliographystyle{ar-style4.bst}

\newpage

\end{document}